\documentclass[fleqn,twoside]{article}
\usepackage{espcrc2}
\usepackage{epsfig}

\usepackage{graphicx}
\usepackage[figuresright]{rotating}

\def\beq{\begin{equation}}
\def\eeq{\end{equation}}
\def\bea{\begin{eqnarray}}
\def\eea{\end{eqnarray}}
\def\bq{\begin{quote}}
\def\eq{\end{quote}}
\def\t{{\mbox{\rm Tr}}}
\newlength{\bredde}
\def\slash#1{\settowidth{\bredde}{$#1$}\ifmmode\,\raisebox{.15ex}{/}
\hspace*{-\bredde} #1\else$\,\raisebox{.15ex}{/}\hspace*{-\bredde} #1$\fi}

\def\gappeq{\mathrel{\rlap {\raise.5ex\hbox{$>$}}
{\lower.5ex\hbox{$\sim$}}}}

\def\lappeq{\mathrel{\rlap{\raise.5ex\hbox{$<$}}
{\lower.5ex\hbox{$\sim$}}}}

\def\Toprel#1\over#2{\mathrel{\mathop{#2}\limits^{#1}}}

\newcommand{\AmS}{{\protect\the\textfont2
  A\kern-.1667em\lower.5ex\hbox{M}\kern-.125emS}}

\title{\vspace{-4.0cm}
Quenched and Unquenched Chiral Perturbation Theory in the $\epsilon$-Regime}

\author{P.H. Damgaard\address{Niels Bohr
Institute, Blegdamsvej 17, DK-2100 Copenhagen, Denmark}}

\begin{document}

\begin{abstract}
The chiral limit of finite-volume QCD is the $\epsilon$-regime
of the theory. We discuss how this regime can be used for 
determining low-energy
observables of QCD by means of comparisons between lattice simulations
and quenched and unquenched chiral perturbation theory. The quenched
theory suffers in the $\epsilon$-regime from ``quenched finite volume
logs'', the finite-volume analogs of quenched chiral logs.

\vspace{1pc}
\end{abstract}

\maketitle
When the chiral limit $m \to 0$ of QCD is taken at finite four-volume $V$, one
will eventually, no matter how big $V$ is, reach a regime where the
pion Compton wavelength $1/m_{\pi}$ exceeds the linear size $L \equiv
V^{1/4}$ of the box. If the volume at the same time is much bigger than
the QCD scale, one is in an interesting region, the $\epsilon$-regime
of QCD, where the spontaneously broken
chiral symmetry is dictating the interactions of the pions, while,
simultaneously, these pions do not even fit inside the box \cite{GL,N}.
It is 
important that there is a huge separation of scales here: Seen from
the mass scale of the pseudo-Goldstone bosons $m_{\pi}$, the rest of hadronic
physics is effectively occuring near the cut-off. This isolates a neat
little corner of QCD which can be studied by analytical means, and the
results are completely non-perturbative in the strong coupling constant.
Far from being an academic playground, this $\epsilon$-regime can thus be
very useful for lattice gauge theory. It is a new regime in which to do
numerical computations in lattice QCD, and where it is extremely important
to get a precise handle on the lowest-lying Dirac eigenvalues \cite{GHLW}.
These smallest eigenvalues dominate physical low-energy
observables in this regime because the mass is taken to zero.

The term $\epsilon$-expansion is used to denote a re-ordered chiral 
perturbation theory. Normally, for volumes infinite or very large compared 
with the pion Compton wavelength, the expansion parameter is basically
``momentum'', $i.e.$ $p^2 \sim m_{\pi}^2$ (hence the alternative name
``$p$-expansion''). In the $\epsilon$-regime
such an expansion is doomed to fail due to the enhanced role played
by momentum zero-modes. For these zero-modes the pion propagator would
blow up, were it not for the mass term $m_{\pi}^2$. As the chiral
limit is taken at fixed four-volume, even this term will blow up. Because
the problem only concerns zero-modes of momentum, a cure is fortunately
available \cite{GL}. The idea is to do the path integral of the 
zero-momentum modes exactly, and treating the remaining modes perturbatively.
This is possible, since the non-zero modes will have a gap on the order
of $1/L$ down to the zero modes. The small quantity $\epsilon$
in which one is expanding is precisely $1/L$. There is a fascinating
relation between the pertinent zero-mode integral of the chiral Lagrangian
and chiral Random Matrix Theory. Much of the recent progress
in understanding the $\epsilon$-regime of QCD stems from this relation
to universal Random Matrix Theory results \cite{V,OTV}. 

While this $\epsilon$-expansion works well for full QCD in the chiral
limit \cite{H}, fully quenched analogs of chiral perturbation theory
\cite{BG,DS} suffer also in the $\epsilon$-regime from the double-pole 
effects of the singlet field \cite{D01,DDHJ}. The probably most drastic
effect of quenching occurs already at the one-loop correction to the
quenched chiral condensate. Just as quenched chiral perturbation theory
in the $p$-expansion shows internal inconsistency by generating 
arbitrarily large corrections in the form of quenched chiral logs
\cite{BG,S}, quenched chiral perturbation theory in the 
$\epsilon$-expansion produces ``quenched finite volume logs'' \cite{D01}.
Although the $\epsilon$-expansion should become better and better as
the volume $V$ is increased, a logarithmic correction prevents this
limit from being taken! The origin of this trouble lies in the fact that
the quenched $\epsilon$-expansion has one additional expansion parameter,
$m_0$, the mass term of the flavor singlet. Contrary to $1/L$, this
parameter is fixed by physics, and cannot be tuned. It means that
the correction term formally can become larger than the leading term,
indicating that the whole expansion, at least as done in this
straightforward way, is sick. The extent to which
one can ignore this problem of quenched QCD will be discussed below. 
Fortunately full and, to some extent, partially quenched QCD does not
have this difficulty, and there the $\epsilon$-expansion converges
nicely as $V \to \infty$.

For simplicity partially or fully quenched chiral perturbation theory will 
be discussed in the framework of the replica formalism \cite{DS,DDHJ}.
All results reviewed here have actually been calculated in the 
supersymmetric formulation \cite{BG} as well, as a check. With replicas,
the full or partially quenched limit in chiral perturbation theory is
quite simple, and differs only in the fully quenched case 
from ordinary chiral perturbation theory, and then only
in the necessity of including the flavor-singlet field together with
the usual Goldstone bosons. The flavor singlet cannot be removed in the
fully quenched theory because there is no replica limit $N\to 0$ of
the flavor group SU($N$). 
As in the supersymmetric formulation, one ``regularizes''
by adding a massive singlet field $\Phi_0(x)$, whose mass $m_0$ in the
replica formulation serves as a deformation of the symmetry group 
away from SU($N$). The SU($N$) limit is reached as $m_0 \to\infty$, 
but that is a limit that cannot be taken before $N\to 0$. The chiral 
Lagrangian then reads 
\begin{eqnarray}
 {\cal L}\! &=& \! {\rm Re}\,
 {\rm Tr} \left[\frac{F_{\pi}^2}{4}\partial_\mu U\partial_\mu U^\dagger - 
 \Sigma\,
 U {\cal M} e^{i\theta/N}
 \right] \cr
&+& {m_0^2 \over 2 N_c} \Phi^2_0 + \frac{\alpha}{2N_{c}}\partial_{\mu}
 \Phi_0 \partial_{\mu}\Phi_0 + \ldots~\label{Lchi}
\end{eqnarray}
Here the ellipses denote terms that are sub-leading in both the
$p$ and $\epsilon$ expansions.
The $\alpha$-term in (\ref{Lchi}) is often omitted (because it has
a power of $p^2$ compared to the $m_0$-term), 
but can in any case easily be
reinstated in all formulas in momentum space by $m_0^2 \to m_0^2
+\alpha p^2$. The constant $\Sigma$ is the (infinite volume) chiral 
condensate, and $F_{\pi}$ is the pion decay constant. The vacuum angle $\theta$
is introduced only in order to be able to project on sectors of
fixed topological charge $\nu$ (see the second of ref. \cite{GL}).

When ordinary chiral perturbation theory based on (\ref{Lchi}) breaks 
down in the chiral limit at finite volume it happens because the zero
momentum modes cannot be treated perturbatively. The cure is to treat
just these modes exactly. In practice this is done by factoring 
\beq
U(x) ~=~ U_0\exp[\sqrt{2}i\xi(x)/F_{\pi}]
\eeq
where $U_0$ is the zero-mode part, and all flutuation fields are included
in $\xi(x)$. In conventional chiral perturbation theory it suffices to
expand also $U_0$ around the origin; here it must be allowed to fluctuate
all over the Goldstone manifold, and the integral over $U_0$ will 
be evaluated exactly. The full integral to be performed is not just over the
action (\ref{Lchi}), but with the observables inserted in the path
integral as well. In practice this is done by evaluating a suitable
generating function of the zero modes exactly.  

\noindent
{\bf The quenched finite-volume logarithm}

Consider the (mass-dependent) quenched chiral condensate
\beq
\Sigma_{\nu}(m) = \lim_{N\to 0}
\frac{1}{NV}\frac{\partial}{\partial m}\ln {\cal Z}_{\nu} ~,
\eeq
where ${\cal Z}_{\nu}$ is the partition function in a sector of
topological charge $\nu$. To leading order in the $\epsilon$-expansion
this can be evaluated analytically on the basis of the zero
momentum integral over $U_0$ \cite{OTV},
\beq
\frac{\Sigma_{\nu}}{\Sigma} = 
\mu(I_\nu (\mu)K_\nu(\mu) + I_{\nu +1}(\mu)
K_{\nu -1}(\mu)) +\frac{\nu}{\mu} ~,\label{sigmanu}
\eeq
where $\mu \equiv m\Sigma V$, and $I$ and $K$ are modified
Bessel functions. This leading order expression has been compared to
lattice gauge theory data with quite nice agreement 
\cite{V1}. But an unpleasant result emerges when we evaluate
the one-loop correction to this result. In the quenched
$\epsilon$-expansion this can be done by computing
the one-loop $\xi$-integral saturation of
$$
\lim_{N\to 0}\frac{1}{N}\left\langle\frac{m\Sigma}{2F^2}{\mbox{\rm Tr}}
\left[(U_0 + U_0^{\dagger})\int\!dx~\xi(x)^2\right]
\right\rangle_{\xi} \label{Nto0}
$$
with, in the replica formalism, $N$ replica quarks of equal mass $m$
(the quenched pseudo-Goldstone bosons thus have common mass
$M = 2m\Sigma/F^2$). To evaluate it, we need the propagators of,
in the quark basis, the off-diagonal mesons $\Phi_{ij} \sim
\bar{\psi}_i\psi_j, i \neq j$,
\beq
D_{ij}(p^2) ~=~ \frac{1}{p^2+M^2} ~, 
\eeq
and the more complicated propagator for the diagonal combination
$\Phi_{ii} \sim \bar{\psi}_i\psi_i$,
\beq
G_{ij}(p^2)\!=\! 
\frac{\delta_{ij}}{p^2+M^2} - \frac{(m_0^2+\alpha p^2)/N_c}{
(p^2 + M^2)^2{\cal F}(p^2)} ~,
\eeq
where \cite{DS}
\beq
{\cal F}(p^2) \!\equiv\! 1 + \frac{(m_0^2+\alpha p^2)N}{N_c
(p^2+M^2)} ~,
\eeq
(and hence equals unity in the quenched $N \to 0$ limit).
Because the integration over $\xi$ excludes zero modes, it is
convenient to define
\beq
\bar{\Delta}(x) \equiv \sum_{p\neq 0} \frac{e^{ipx}}{p^2}~~;~~
\bar{G}(x) \equiv \sum_{p\neq 0} \frac{e^{ipx}}{(p^2)^2}
\eeq
and note that these are just obtained from the second and third terms in the
Taylor expansion of the pion propagator at $x=0$:
\beq
\frac{1}{V}\sum_{p} \frac{e^{ipx}}{p^2+M^2} = \frac{1}{M^2V}
+ \bar{\Delta}(x) - M^2\bar{G}(x) ~,\label{Taylorprop}
\eeq
in an obvious notation. Performing the contractions, 
one gets, successively,
\bea
&& \lim_{N\to 0}\frac{1}{N}
\frac{Vm\Sigma}{2F_{\pi}^2}\t(U_0 + U_0^{\dagger})\times \cr
&& \left\{(N-1)\bar{\Delta}(M^2) + \frac{1}{V}\sum_{p\neq 0}
G(M^2)\right\}_{N\to 0}\cr
&=& \lim_{N\to 0}\frac{1}{N}
\frac{Vm\Sigma}{2F_{\pi}^2}\t(U_0 + U_0^{\dagger})\times \cr
&& \left\{(N-1)\bar{\Delta}(M^2) + \bar{\Delta}(M^2)\right.\cr
&& - \left.\frac{1}{V}\sum_{p \neq 0}\frac{\frac{1}{N_{c}}
(m_0^2+\alpha p^2)}{(p^2+M^2)^2{\cal F}(p^2)}\right\}_{N\to 0} \cr
&=&  - \lim_{N\to 0}\frac{1}{N}\frac{Vm\Sigma}{2F^2}\t(U_0 + 
U_0^{\dagger})\times \cr
&&
\frac{1}{V}\sum_{p\neq 0} \frac{\frac{1}{N_{c}}
(m_0^2+\alpha p^2)}{(p^2+M^2)^2} ~.
\eea
This correction can be rewritten
in precisely the same form
as the original mass term in the Lagrangian. We can thus re-exponentiate,
and find that to this order in the quenched $\epsilon$-expansion we
can use the same effective Lagrangian, but with a shifted mass term.
In detail, we replace $\mu \equiv m\Sigma V$ by $\mu'$, where
\beq
\frac{\mu'}{\mu} \equiv \left\{1 + \frac{1}{N_{c}F_{\pi}^2}\left[
m_0^2\bar{G}(0)+ \alpha\bar{\Delta}(0) \right]\right\} 
\eeq
and compute the one-loop improved chiral condensate from
\beq
\frac{\Sigma(\mu)}{\Sigma} ~\equiv~ \lim_{N\to 0}
\frac{1}{N}\frac{\partial}{\partial \mu'}
\ln {\cal Z}(\mu')\cdot\frac{\mu'}{\mu} ~. \label{qconddef'}
\eeq
Alternatively, to this order one is effectively working with a
shifted $\Sigma$-parameter in finite volume. To bring out the
volume dependence it pays to rewrite it as
\beq
\frac{\Sigma_{eff}}{\Sigma} = 1 + \frac{1}{N_cF^2}[m_0^2\bar{G}(0)
+ \alpha\bar{\Delta}(0)] ~,\label{sigmaeff}
\eeq
valid to this order and in the chiral limit. In dimensional regularization
$\bar{\Delta}(0)$ is finite \cite{HL}, 
\beq
\bar{\Delta}(0) ~=~ -\beta_1/L^2 ~, \label{deltabar}
\eeq
where $\beta_1$ is a universal ``shape coefficient'' that depends
only on the geometry of the finite volume.
There is, even in dimensional regularization,
an ultaviolet divergence associated with $\bar{G}(0)$,
\beq
\bar{G}(0) ~=~ \beta_2 + \frac{1}{8\pi^2}(\ln(L\Lambda_0) - c_1) ~,
\label{gbar}
\eeq
where $\beta_2$ is another universal shape coefficient \cite{HL},
$\Lambda_0$ is the momentum space subtraction point,
and $c_1$ is, still in dimensional regularization,
\beq
c_1 = \frac{1}{d-4} - \frac{1}{2}(\Gamma'(1) + 1 + \ln(4\pi) +
{\cal O}(d-4)) ~.
\eeq
In fact, this divergence found here in the $\epsilon$-expansion
matches precisely the new quenched one-loop counterterm of the
infinite-volume theory found by Colangelo and Pallante \cite{CP}.
This is as it should be: the divergence structure at finite
volume is just as in infinite volume, no new counterterms
are needed, and the already existing infinite-volume
counterterms are cancelled by similar divergences in the
finite-volume theory. 

If we plug (\ref{deltabar}) and (\ref{gbar}) into the expression
for the effective chiral condensate parameter $\Sigma_{eff}$ in
(\ref{sigmaeff}) we find a logarithmic divergence in $L$, the
{\em quenched finite volume logarithm}. As this is supposed to be an
expansion in $1/L$, something has obviously gone terribly wrong.
It is worth noting that the logarithmic term (\ref{gbar}) occurs
even in the unquenched theory, but there always accompanied by
inverse powers of $L$, which makes its appearance there innocuous.
We trace the origin of the logarithmic disaster in the quenched
theory back to the strong infrared divergence associated with
a double-pole propagator. This failure of the quenched 
$\epsilon$-expansion occurs because there really is one more
expansion parameter involved, $m_0$, which is not a tunable
quantity. Rather $m_0$ is related to the flavor singlet mass of
the unquenched theory, and it remains a finite number even in
the chiral limit.
The quenched finite-volume logarithm occurs in the two other
chiral symmetry breaking classes too \cite{Jesper}, and thus
appears to be unavoidable consequence of quenching. The
fact that what should be a small correction to the leading-order
result can become big means that the expansion as
it stands cannot be trusted. To this order, the best one can do
is to view it as a prediction for the finite-volume scaling
behavior of nearby volumes $V_1$ and $V_2$. Then the perturbation theory to
this order still makes sense, and one gets the prediction
\bea
\frac{\Sigma_{eff}(V_1)}{\Sigma_{eff}(V_2)} &=& 1
-\frac{1}{N_cF_{\pi}^2}\left[\alpha\beta_1\left(\frac{1}{L_1^2}-\frac{1}{L_2^2}
\right)\right. \cr
&&\left.-\frac{m_0^2}{8\pi^2}\ln\left(\frac{L_1}{L_2}\right)\right]
\eea
to this order. Taken at face value it suggests that the infinite
volume chiral condensate one attempts to extract from the quenched
theory may keep growing as the volume is increased. This type
of behavior has often been suspected, and there is also
numerical evidence from two space-time dimensions to support it \cite{KN}.

\noindent
{\bf Quenched Correlation Functions}

The $\epsilon$-expansion gives rise to a new chiral perturbation
theory for the low-energy observables, and in particular $n$-point
functions of mesonic operators can be worked out in this framework.
The quenched finite volume logarithms, and quenching artifacts
in general, will put restrictions on the applicability of these
expressions in the quenched theory. Nevertheless, since this is where
the main numerical focus is with present-day techniques for fermion
simulations with good chiral properties, it is of interest to
determine these quenched expressions. So far computations have
been performed for two-point functions of both scalars (S) and
psuedoscalars (P) \cite{DDHJ} as well as vectors (V) and axial vectors
(A) \cite{DHJLL}. Also certain quenched three-point functions
have been computed \cite{PM}. Until this year only one preliminary
numerical study using domain wall fermions had explored these
quenched correlation functions in the $\epsilon$-regime \cite{PO}, but 
one month ago new overlap-fermion results were reported
\cite{DESY}.

For the flavor singlet correlation functions there is no need to
work with anything but one valence quark, $N_v=1$. The relevant
operators at quark level read, in a replica notation,
\bea
S^0(x) &\equiv& \bar{\psi}(x)I_{N_{v}}\psi(x) \cr
P^0(x) &\equiv& \bar{\psi}i\gamma^5I_{N_{v}}\psi(x) ~,
\eea
where the $N\times N$ matrix $I_{N_{v}}$ is unity in the first
entry, and zero elsewhere. The quenched limit is taken by sending
$N\to 0$ \cite{DDHJ}. Sources $s(x)$ and $p(x)$
for such operators are readily transcribed into
the effective Lagrangian by means of replacing the source (mass matrix)
\beq
{\cal M} \to {\cal M} +s(x)I_{N_{v}} + ip(x)I_{N_{v}}
\eeq
in the chiral Lagrangian (\ref{Lchi}). Flavored sources are added
analogously: we now take $N_{v}$ to be (at least) 2, define 
$I_{N_{v}}$ correspondingly, and then transcribe the quark
currents
\bea
S^a(x) &\equiv& \bar{\psi}(x)t^aI_{N_{v}}\psi(x) \cr
P^a(x) &\equiv& \bar{\psi}it^a\gamma^5I_{N_{v}}\psi(x) 
\eea
into the effective Lagrangian by means of the sources
\beq
{\cal M} \to {\cal M} +s^a(x)t^aI_{N_{v}} + ip^a(x)t^aI_{N_{v}} ~.
\eeq
Similarly for the vector and axial vector currents which at the quark level
read
\begin{eqnarray}
V^a_\mu(x) &\equiv& \bar{\psi}(x) i \gamma_\mu t^aI_{N_v} \psi(x)  \cr
A^a_\mu(x) &\equiv& \bar{\psi}(x) i \gamma_\mu\gamma_5t^a I_{N_v} \psi(x) ~.
\end{eqnarray}
and which can be assigned sources in the effective theory through
a replacement of $\partial_{\mu}U(x)$ by a covariant derivative,
as in ordinary chiral perturbation theory. In the 
$\epsilon$-expansion the pertinent correlation functions are
then evaluated by performing the loop epxansion in the fluctuating
$\xi(x)$, while performing the zero-mode integral over $U_0$
exactly. It is this last step which is the tricky part. Fortunately
all analytical results required for $N_v=2$ compuations are
available in the literature \cite{OTV}, after suitable
manipulations \cite{DDHJ}.

The analytical results for these two-point functions are fairly
simple in structure, and all depend on two kinematical functions
\cite{HL}
\bea
h_1(\tau) &=& \frac{1}{T}\int\!d^3x~\bar{\Delta}(x)\cr
&=& \frac{1}{2}\left[(\tau-\frac{1}{2})^2-\frac{1}{12}\right]\cr
h_2(\tau)&=& - \int\!d^3x~\bar{G}(x) \cr
&=& \frac{1}{24}\left[\tau^2(1-\tau)^2-\frac{1}{30}\right] ~,
\eea 
where $T$ is the extent in the ``time'' direction and
$\tau \equiv t/T$. These two functions replace the simple 
$\cosh$-function of the usual two-point functions in conventional
chiral perturbation theory.

To give an example of the kind of expression one gets, consider
the {\em sum} \cite{DDHJ}
\bea
&& \langle S^a(x)S^a(0)\rangle + \langle P^a(x)P^a(0)\rangle = A(\mu)+\cr
&& \frac{\Sigma}{F_{\pi}}\left[\bar{\Delta}(x)-B(\mu)(m_0^2\bar{G}(x)
+ \alpha\bar{\Delta}(x))\right] ~,
\eea
where $A(\mu)$ and $B(\mu)$ are known explicit functions of $\mu$:
\bea
\frac{A(\mu)}{\Sigma} &=& \frac{1}{2}\left[\Sigma_{\nu}^{1-loop'}(\mu)
+\frac{1}{\mu}\Sigma_{\nu}^{1-loop}(\mu)\right] \cr
B(\mu) &=& \frac{1}{N_c}\left(\frac{\Sigma_{\nu}'(\mu)}{\Sigma}
+\frac{1}{\mu}\frac{\Sigma_{\nu}(\mu)}{\Sigma}\right) ~.
\eea
The function $\Sigma_{\nu}(\mu)$ is as given in eq. (\ref{sigmanu}),
and the ``1-loop improvement'' is as dictated by the 1-loop correction
to the chiral condensate as discussed above,
\beq
\Sigma^{1-loop}_{\nu}(\mu) ~\equiv \Sigma_{\nu}(\mu')\frac{\mu'}{\mu} ~.
\eeq
Taking, as usual on the lattice, the projection onto zero momentum
of this sum of correlation functions is then straightforward and
expresses the spatial integral of this sum in terms of the kinematical
functions $h_1(\tau)$ and $h_2(\tau)$. Although this looks quite
different from conventional two-point functions on the lattice, one
can check explicitly that the above result matches the expression
from ordinary chiral perturbation theory in the region of overlap.
The relation (\ref{Taylorprop}) is crucial in establishing this.

We have already mentioned that in the $\epsilon$-regime there is
strong dependence on gauge field topology. This is clearly seen
in the correlation functions which vary strongly with increasing
$\nu$. Exceptions are suitable combinations of correlation functions
such as the one discussed above, where the quenched poles at zero
quark mass in non-trivial gauge field backgrounds cancel at any
fixed $\nu \neq 0$. Zero topology, $\nu =0$, is yet again special:
there is no quark mass pole but still a strong infrared sensitivity
to the quark mass, see ref. \cite{DDHJ} for details.

Even when explicit power-like divergences in the inverse quark mass
are absent, these quenched correlation functions all suffer from
the quenched finite volume logs. This is clear from the example
$\langle S^a(x)S^a(0) + P^a(x)P^a(0)\rangle$ discussed above since
the 1-loop improved chiral condensate $\Sigma^{1-loop}(\mu)$ enters
directly in this correlation function. Again, taken at face value
the analytical prediction is a divergence in the correlation
function. But the divergence is also here not to be trusted, since
it arises as a {\em correction} to the leading order result. As for
the chiral condensate itself one can hope that the analytical prediction
gives the scaling between two nearby volumes, where no divergences
arise. It should also be stressed that these quenched finite
logs are by now means worse than the more familiar quenched chiral
logs; only the fact that in the $\epsilon$-regime one really
aims at taking the chiral limit makes the issue of these quenched
infrared divergences more urgent to resolve.

If one computes vector and axial vector correlation functions, the
quenched finite volume logs do not appear neither at leading nor
subleading order in the $\epsilon$-expansion \cite{DHJLL}. This
should not be taken as evidence that these correlation functions
are free of these logs, only that they do not appear at these
first 2 non-trivial orders (corrections to the quenched chiral condensate,
and hence quenched finite logs are almost bound to appear eventually).
Actually one of these, the quenched vector-vector correlation is
free of these logs, but in a trivial way:
\beq
\langle V_0^a(x)V_0^a(0)\rangle = 0
\eeq
to all orders in the quenched $\epsilon$-expansion \cite{DHJLL}! The
proof of this statement can be found in ref. \cite{DHJLL} (it is
a peculiar quenched artifact with a simple diagrammatic interpretation).
The 2-point function of axial vectors also shows a quite remarkable
behavior. Up to and including next-to-leading order,
\bea 
\int\!\!d^3x\langle 
 {A}^a_0(x) {A}^a_0(0) \rangle\!\! 
 = \!\! - \!\frac{2F_{\pi}^2}{T}\!\left(\! 1\! +\! 
\frac{4 m \Sigma_\nu(\mu) T^2}{F_{\pi}^2} 
 h_1(\tau)\!\!\right)
\eea
This result comes about after a series of non-trivial cancellations
where all double-pole terms in the quenched $\epsilon$-expansion
have cancelled each other. To leading order this gives a very simple
prediction that does not refer to quenched parameters at all: the
axial vector 2-point function directly measures the pion decay
constant $F_{\pi}$. Including next-to-leading order it also determines
the quenched chiral condensate (but actual numerical simulations
may have difficulty disentangling this correction from the leading
result \cite{DESY}).

\noindent
{\bf Predictions for full QCD:}

The difficulties of the quenched $\epsilon$-expansion are fortunately
absent in the full theory. The quenched finite volums logs become,
in the full theory, multiplied by inverse powers of the volume, and
are hence rendered harmless. This is entirely analogous to the
quenched chiral logs becoming replaced by ordinary  (harmless)
chiral logs in full QCD in the $p$-expansion. Just as ordinary
chiral logs are multiplied by powers of the pion mass, the finite
volume logs in the $\epsilon$-expansion become multiplied by inverse
powers of the volume.

Analytical predictions for full QCD
in the $\epsilon$-regime was worked out long ago by Hansen \cite{H},
and they have recently been re-done \cite{DDHJ,DHJLL} 
for sectors of fixed gauge field topology. These predictions in
sector of fixed topology may
eventually may be of most interest from a numerical point of view,
in particular since there are distinct predictions for each
particular $\nu$-sector.

Also in the full theory it turns out that all zero-mode integrals
can be expressed in terms of the mass-dependent chiral
condensate $\Sigma_{\nu}(\mu)$
and derivatives thereof. As in the quenched theory it pays to define,
to one-loop order a shifted mass which here turns out to be \cite{GL}
\begin{eqnarray}
\frac{\mu'}{\mu} ~\equiv~ 1 + \frac{N_f^2-1}{N_f}\frac{\beta_1}
{F^2\sqrt{V}} ~.
\end{eqnarray}
All analytical predictions are then expressed in terms of the
kinematical functions $h_1(\tau)$ and $h_2(\tau)$ up to an including
next-to-leading order in the $\epsilon$-expansion (additional
numerical factors related to the lattice volume also appear).
Details can be found in refs \cite{DDHJ,DHJLL}.

{\sc Acknowledgements:}~ I thank M.C. Diamantini, P. Hernandez, 
K. Jansen, M. Laine, L. Lellouch and K. Splittorff for enjoyable
collaborations on the subject described here.

\end{document}